# An atomic perspective on the serpentine-chlorite solid-state transformation

Hui Zhang[1], Benjamin Gilbert[1], Jillian F Banfield[1,2]*

[1]Energy Geoscience Division, Lawrence Berkeley National Laboratory, Berkeley, CA 94720 USA

[2]Department of Earth and Planetary Science, University of California, Berkeley, CA 94720 USA

**Abstract**

**Serpentine minerals are important components of metamorphic rocks and promising geo-materials for nanotechnology. Lizardite, the most abundant serpentine mineral, can be transformed into chlorite during metamorphism. This intriguing phase transformation should affect the deformation behavior during aseismic creep and slow slip at the base of the subduction zone, but has not been understood structurally and chemically at the atomic scale. Here we visualized cations and oxygen atoms using the state-of-the-art low-dose scanning transmission electron microscopy and found that restructuring mainly involves the synergistic migration of tetrahedral cations and oxygen anions, coupled with the migration of octahedral trivalent cations into the brucite-like interlayer. Further, we show that different serpentine polytypes result in distinct regular interstratifications of serpentine and chlorite. Our results clarify the long-standing puzzle of how solid-state layer silicate transformations occur and lead to long-period ordered structures.**

* Corresponding author.
Email address: jbanfield@berkeley.edu

## Introduction

Clay minerals are abundant in the Earth's crust and have been used for nanotechnology[1] and to trace the seismic[2-4] and climatic[5] history of the earth. Clay minerals are architectured by sheets of metal-oxide tetrahedra (T) and octahedra (O) with magnesium, aluminum and silicon as the common cations.[6] The geometrical assembly of TO layers at the atomic scale creates a diversity of minerals, the structurally simplest family of which is serpentine. Most serpentine minerals are Mg-rich phyllosilicates composed of one T-sheet and one O-sheet with all of the octahedral sites occupied. Because of their layered structures and surface chemical properties, these minerals have been used for civil construction, toxic metal absorption, $CO_2$ capture and as catalyst carriers.[7,8]

Serpentine minerals play vital roles in the seismicity of the subduction zone[2-4,9,10] because serpentine minerals and rocks exhibit anisotropic mechanical properties that can enable aseismic creep[11-13]. Serpentine minerals can transform to chlorite[14-21], a layer silicate with greater compressibility anisotropy[22,23], a process that is expected to enhance seismicity at intermediate depths. The pathway of this solid-state transformation of serpentine to chlorite remains only partially understood. Based on high-resolution transmission electron microscopy (HRTEM) analysis of lizardite[18,21] and a related mineral berthierine[19], the transformation essentially requires the T-sheet of one TO layer to flip and combine with another TO layer, forming a TOT layer and transformation of the O-sheet of the former TO layer to a separate brucite-like (B) interlayer.[18,21]

Fundamental structural details regarding the lizardite precursor and chlorite product remain debated. Specifically, one prior HRTEM study proposed that the transformation of serpentine polytypes involves shifts of the inverted T-sheet and O-sheet in the newly-formed TOT layer that preserve the optimal hydrogen bonding between the TOT surface and the B-interlayer.[18] In contrast, another HRTEM study proposed that neither the serpentinite nor the chlorite adopted the hydrogen-bonded configuration.[21] Because interlayer hydrogen bonding is believed to be an important energetic contribution to layer silicate stability, establishing the role for this interfacial bonding in this important transformation has wide-ranging implications for all clay minerals. In addition, lizardite and chlorite frequently exhibit polytypism, yet the structural and energetic controls on polytype formation are very poorly understood and the relationship between the precursor and product phases is unknown. More generally, the overall driving force for this transformation remains poorly defined and not all chemical and structural contributions have been established. A limitation of prior structural studies was that the image resolution was not sufficient to directly determine the position of oxygen atoms at the transformation interface, and hence the crystallographic shifts could not be unambiguously determined. Further, the prior studies could not visualize any associated redistribution of octahedral cations.

Here, we used the low-dose HAADF–STEM method to elucidate the atomic-scale details of the serpentine to chlorite transformation process in lizardite, the most abundant serpentine mineral. Unlike conventional HRTEM, high-angle annular dark-field (HAADF) scanning transmission electron microscopy (STEM) can directly image the atomic columns of clay minerals and the image contrast is straightforwardly interpretable and in proportion to $\overline{Z^\alpha}$ (the average of $Z^\alpha$ for all the atoms along the projection direction), where Z is the atomic number and the exponent $\alpha$=1–2.[24,25] Atomic resolution has been achieved in HAADF–STEM images of minerals like mica,[26] Cs-bearing phlogopite,[27] Fe-rich celadonite[28] and Fe-rich berthierine[29]. Some hydrous minerals that lack Fe are generally very electron-beam sensitive. For example, damage of the aluminosilicate clay mineral halloysite starts and completes at an electron dose of ~1950 and ~8800 $e^-/Å^2$. Doses of less than ~100 e⁻/Å² were required to obtain a [001] image of the hydrous aluminosilicate montmorillonite with a 1.2 Å resolution.[30] Serpentine minerals are less beam sensitive than halloysite and montmorillonite, so they are amenable to low-dose HAADF-STEM, which forms images using only a small fraction of the electrons that are scattered to high angles. Recently, we achieved 1.0 Å resolution in HAADF–STEM images of serpentine polytypes with doses of ~6000 e⁻/Å².[31]

Here, we report the results of a high-resolution study that clearly resolved the structural and chemical details of the transformation pathway that leads to chlorite in the hydrogen-bonded configuration. We provide a model for the transformation mechanism that is applicable to all lizardite precursor polytypes that is based on the direct observation that the tilts of the cation-oxygen octahedra are always preserved throughout the transformation. This model further illustrates how chlorite polytypes are inherited from the stacking structure of the precursor lizardite polytypes, confirming the previously proposed link between the polytype of the serpentine precursor and the pattern of serpentine-chlorite interstratification in the reaction product.[18] The sensitivity of HAADF–STEM to atomic number revealed a chemical redistribution that is clearly coupled with the T-layer inversion, as transition metal cations that are initially randomly distributed in lizardite O-sheets are observed to partition completely to the B-interlayer in chlorite. This approach could also be used to the study of atomic-scale phase transformation mechanism involving other minerals.[16,32,33]

**Experiment**

We studied a sample of Cr-containing lizardite (Cr-lizardite) that is intergrown with chlorite from Woods Chrome Mine in the State Line Serpentinite, Lower Britain Township, Lancaster County, Pennsylvania.[34] The sample was from the same specimen used in a prior study Ref. 31.

Energy dispersive X-ray spectrometry analysis indicates a formula of $(Mg_{0.877}Cr_{0.069}Al_{0.034}Fe_{0.018})_3(Si_{0.827}Al_{0.173})_2O_5(OH)_4$. TEM foils were prepared by ion milling and coated with carbon to mitigate the charging. Electron diffraction patterns and HAADF–STEM images were collected on the Thermo Fisher Themis and double-Cs-corrected TEAM1[35] operating at 300 kV with a probe convergence semi-angle of 17.1 mrad. Since the sample is very beam sensitive, the electron dose was kept around 6000 e$^-$/Å$^2$. To reduce the electron irradiation as much as possible, all the atomic-scale HAADF–STEM images were collected from the region near the area used for sample and microscope alignment, with an inner collection semi-angle of 48 mrad. The images were denoised using the average background subtraction filter in Gatan DigitalMicrograph. The image simulations were performed with a multi-slice algorithm embedded in Prismatic[36] using the experimental microscope parameters. The sample thickness was measured to be ~30 nm by the electron energy loss spectra (EELS) collected on the K3 detector. The models with 30 nm in thickness were sliced into ~1 Å thick slabs and 30 frozen phonon configurations were used to account for the thermal diffuse scattering.

## RESULTS

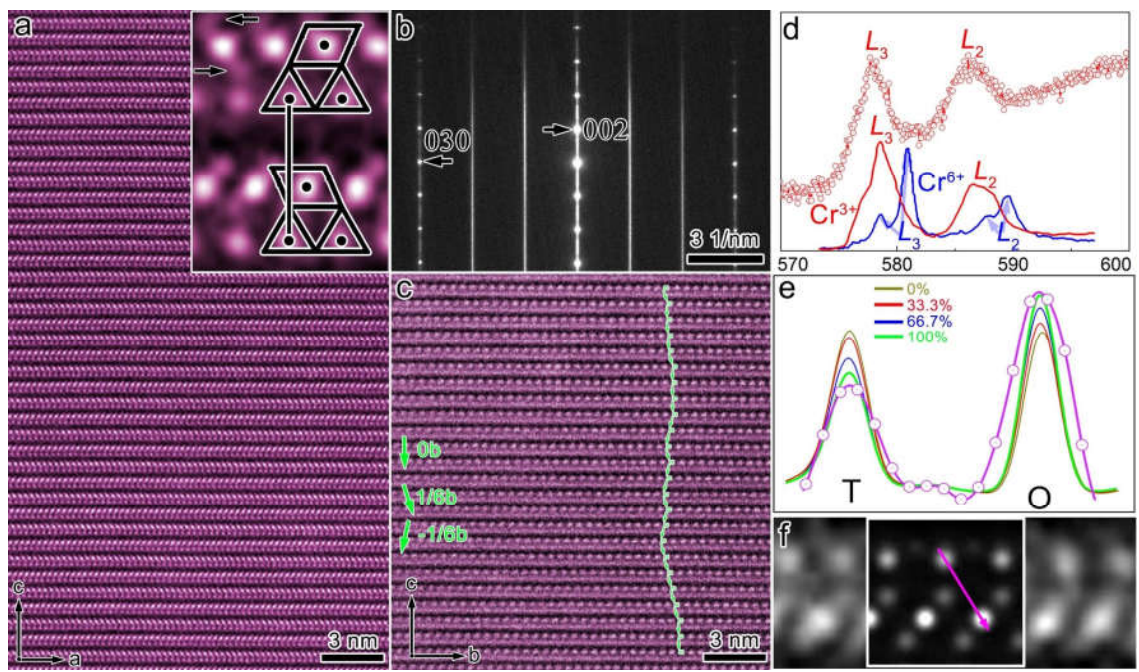

**Fig. 1. Structure of Cr-lizardite.** (a) Low-dose HAADF–STEM image of lizardite layers with viewing direction along [010]. An enlarged image is inserted on the top right, where triangles and parallelograms with dots at the centroid illustrate the tetrahedra and octahedra respectively. Arrows in the inset mark oxygen atomic columns. (b) SAED pattern and (c) HAADF–STEM image along [100]. The SAED pattern was indexed with a pseudo-orthogonal cell with $a$=5.3 Å, $b$=9.2 Å, $c$=14.2 Å, $\alpha$=90.3°, $\beta$=90.1° and $\gamma$=90.0°. In (c), small boxes mark tetrahedral cation pairs and arrows illustrate interlayer displacement along the $b$-axis. 0$b$, 1/6$b$ and -1/6$b$ correspond to an interlayer shift of |0$b$|, |1/3$b$| and |-1/3$b$|. (d) EELS spectra of the Cr $L_{2,3}$-edge. Red and blue lines are $Cr^{3+}$ and $Cr^{6+}$ spectra adapted from Ref. 37. Red line with markers is the spectrum of Cr-lizardite. (e) Intensity profiles of simulated and experimental images along the line indicated by pink arrow in (f). Brown, red, blue and green lines are from the model with 0%, 33.3%, 66.7%, 100% Cr and Fe in octahedra. Pink line with markers is an experimental profile. T and O denote the peak of tetrahedral and octahedral cations, respectively. (f) Composite image with simulated image imposed on experimental image.

**Lizardite Structure and Chemistry**

Low-dose HAADF–STEM imaging of the Cr-lizardite fully revealed the stacking and distribution of transition metal substitutions. As shown in the [010] image (**Fig. 1a**), T-sheets are stacked without any interlayer shifts along the $a$-axis. The position of oxygen atomic columns, marked by arrows in the inset, relative to the octahedral cations indicates that O-sheets in adjacent TO layers tilt alternatively towards positive and negative *a*-axes. Like other clay minerals[20,38,39], the interlayer shifts along the *b*-axis are semi-random, based on the complete streaking of (01*l*) and (02*l*) (**Fig. 1b**) and the irregular interlayer displacement in the [100] image (**Fig. 1c**). Despite this, the selected area electron diffraction (SAED) patterns along [010] (**Fig. 1b**) and [100] (**Fig. S1**) can be both indexed with a pseudo-orthogonal cell with *a*=5.3 Å, *b*=9.2 Å, *c*=14.2 Å. Based on the stacking of T-sheet relative to the O-sheet in adjacent TO layers shown in **Fig. 1a**, the distance between the anion centers in interlayer hydrogen-bonds is ~2.8 Å. The lizardite layers are thus hydrogen-bonded.[21,40]

Cr in minerals has two oxidation states, $Cr^{3+}$ which is generally in octahedral sites, and $Cr^{6+}$ which tends to occupy tetrahedral sites. The $L_3$ and $L_2$ edges for $Cr^{6+}$ in EELS spectra have two separate peaks, but only one broad peak with shoulder peaks is present for $Cr^{3+}$.[37] The EELS spectrum in **Fig. 1d** suggests that Cr in our sample is tri-valent, thus likely located in octahedral sites, as it is in mica, smectite and halloysite.[41,42] Fe was not resolved in EELS due to its very low concentration. To further study the residency of Cr and Fe, HAADF–STEM images were simulated with different Cr and Fe concentrations in octahedra and compared with the experimental images. As shown in **Fig. 1e,f**, the intensity profile of the model with all Cr and Fe in octahedra (green line in **Fig. 1e**) agrees best with the experimental profile (pink line). Given that the intensities of O-sheets are homogenous (**Fig. 1a**), Cr and Fe are thus treated as evenly distributed octahedral cations.

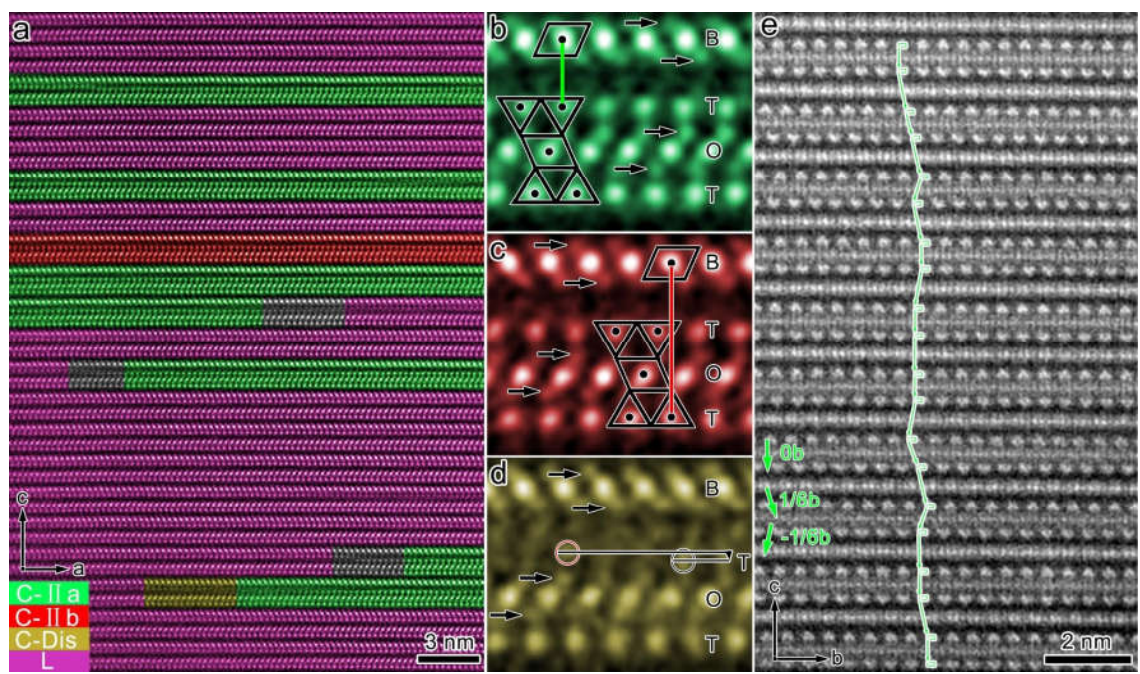

**Fig. 2. Chloritization of lizardite layers**. (a) Low-dose HAADF–STEM image along [010] showing the co-existence of lizardite (L, pink) and chlorite layers separated by transition zones (grey). Three distinct chlorite regions were observed and colored: the chlorite IIa polytype (C–IIa, green), the chlorite IIb polytype (C–IIb, red) and disordered chlorite (C-Dis, brown). Zoomed-in image of C–IIa (b), C–IIb (c) and C-Dis (d). In (b–d), the brucite-like interlayer, octahedral and tetrahedral sheets are denoted by B, O and T, respectively. Arrows mark oxygen atomic columns in the O-sheet. Two atomic columns circled by red and black lines in (d) are ~0.5 Å apart from each other along the *c*-axis. (e) [100] image of chlorite from a fully transformed region of the sample. Small boxes mark tetrahedral cation pairs and arrows illustrate interlayer displacement along the *b*-axis. 0*b*, 1/6*b* and -1/6*b* correspond to interlayer shifts of |0*b*|, |1/3*b*| and |-1/3*b*|.

## Chloritization of Lizardite Layers

The lizardite layers frequently transitioned into chlorite layers (**Fig. 2a**). Two polytypes of chlorite single layers formed in this way were identified as the IIa (C-IIa, green) and IIb (C-IIb, red) as well as disordered structure (C-Dis, brown). For C-IIa (**Fig. 2b**), the octahedral cations in B-interlayer are aligned on the *c*-axis with the upper tetrahedral cations in the TOT layer, but they are aligned on the *c*-axis with the lower tetrahedral cations in the TOT layer in C-IIb (**Fig. 2c**). In C-Dis (**Fig. 2d**), the upper tetrahedral cations in the TOT layer are disordered but some atomic columns are still identifiable. Statistics from over 1500 chlorite layers demonstrate that 73% of single chlorite layers are IIa, while 26% are IIb and 1% are disordered. Like lizardite, the interlayer shifts in chlorite along the *b*-axis are irregular (**Fig. 2e**).

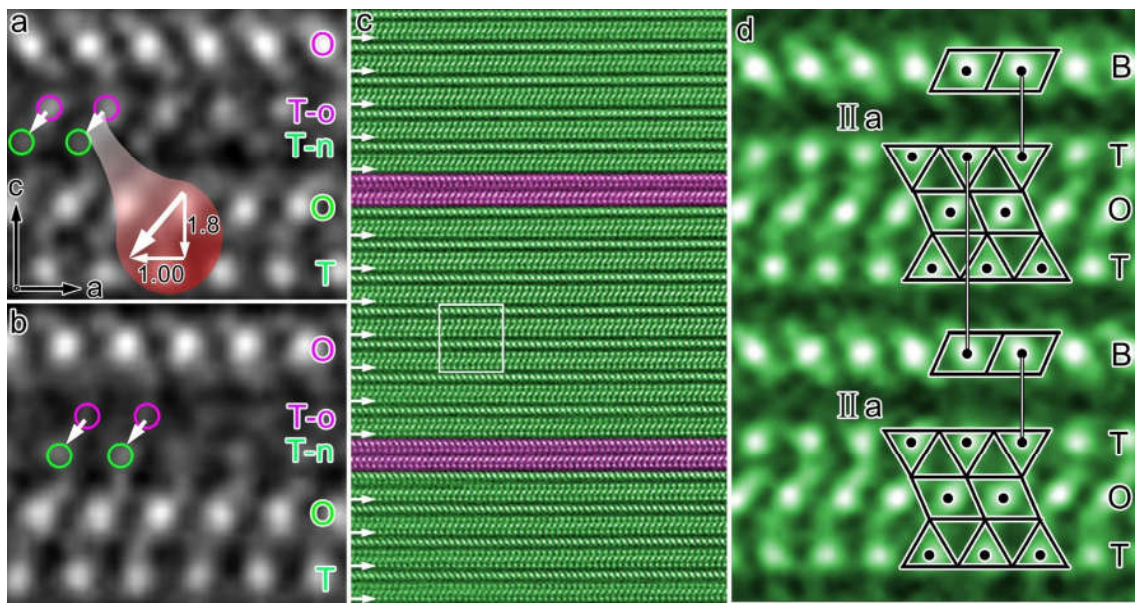

**Fig. 3. Formation of C–IIa chlorite**. (a) The early and (b) late stages of the transformation from lizardite to chlorite. Tetrahedral cations, highlighted by pink circles, migrate to the position marked by green circles. Displacements are illustrated by white arrows, which correspond to 1.8 Å along -*c* and 1.0 Å along -*a*. Brucite-like interlayer, octahedral and tetrahedral sheets are denoted by B, O and T. T-n and T-o denote the new and original T-sheets during the transformation, respectively. (c) C–IIa chlorite (green) intergrown with lizardite (pink). The framed region is enlarged in (d), where triangles and parallelograms with dots at the centroid illustrate tetrahedra and octahedra respectively.

The grey regions in **Fig. 2a** are transition zones, which comprise ~10 to ~20 cation columns along the *a*-axis, and from which we can infer details of the serpentine-chlorite transformation. **Fig. 3a** shows an early transformation stage of lizardite to C-IIa. The tetrahedral cations marked by pink circles in the original T-sheet (T-o) sheet migrate ~1.8 Å along -*c* and ~1.0 Å along -*a* to green circle positions in the new tetrahedral sheet (T-n). The intensity of atomic columns in T-n is about half that of atomic columns in T-o, indicating that ~40% tetrahedral cations have out-migrated from the T-o to T-n sheet using the simple $Z^2$ rule.[24,25] As the transformation proceeds, the atomic columns in the T-n sheet show a higher intensity than those in T-o sheet in the late stage (**Fig. 3b**). In **Fig. 3b**, ~70% tetrahedral cations are now in T-n. When all the T-o cations are in T-n forming a TOT layer and the O-sheet in the upper TO layer is now a B-interlayer, the transformation is complete (**Figs. 3c,d**).

We can unambiguously resolve the orientation of the octahedra to determine the tilt of O-sheets in lizardite and chlorite. Using the untransformed lizardite layers in **Fig. 3c** as a reference, it is apparent that the tilt of the octahedra in chlorite layers is the same as in contiguous lizardite layers. Thus, the tilt of the O-sheet did not change during the phase transformation. Besides, it is observed that the unchanged T-sheets (highlighted by white arrows) in chlorite have no interlayer shift along the *a*-axis, exactly as in lizardite (**Fig. 1a**). As shown in **Fig. 3d**, the uppermost tetrahedral cations in the TOT layer of the upper IIa chlorite layer superimpose on the octahedral cations in the B-interlayer of the lower IIa chlorite layer, suggesting that the adjacent IIa chlorite single layers are stacked in an IIb manner. The overall configuration is IIab,[43] which accounts for 86% in the multiple chlorite layers with the balance being 14% IIbb.

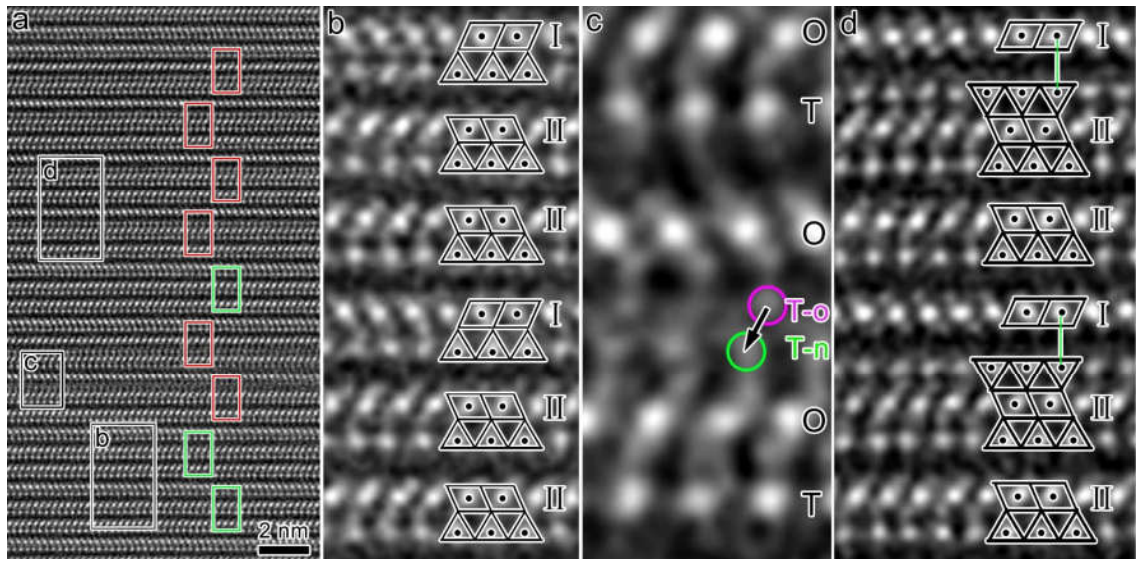

**Fig. 4. Ordered serpentine-chlorite interstratification.** (a) HAADF–STEM image showing the co-existence of transformed (red rectangles) and un-transformed (green rectangles) 3-layer lizardite structure. (b) High-

magnification image of lizardite with an octahedral tilt pattern (I,II,II). (c) Serpentine-chlorite interface region that is partially transformed along the viewing directions. Octahedral and tetrahedral sheets are denoted by O and T. T-n and T-o denote the new and original T-sheets during the transformation, respectively. (d) Ordered serpentine-chlorite interstratification, $S_1C_1$.

The inheritance of octahedral tilts also occurs in the chloritization of long-period lizardite structures, which are significantly less common than the 2-layer lizardite structure[31,34,40], which is based on alternating octahedral tilts, as shown in **Fig. 1a**. **Fig. 4a** shows the co-existence of 3-layer (**Fig. 4b**), partially (**Fig. 4c**) and completely (**Fig. 4d**) chloritized lizardite layers. The 3-layer long-period lizardite has an octahedral tilt pattern (I,II,II). With similar tetrahedral cation migrations from T-o to T-n sheet shown in **Fig. 4c** and **Figs. 3a,b**, two TO layers transform to a chlorite layer. If the complete transformation occurs regularly for every second TO layer, a structure based on the regular alternation of one serpentine (S) and one chlorite (C) layer is created (**Fig. 4d**). This $S_1C_1$ interstratification is the mineral dozyite.[44] In $S_1C_1$, the tilt pattern is fully inherited from 3-layer lizardite structure (**Figs. 4b,d**). The inheritance was also observed in $S_2C_1$ and $S_2C_3$ (**Fig. S2**). In some regions we imaged out-of-phase interfaces, where the octahedral sheets of chlorite layers on one side of the boundary are continuous with brucite-like interlayers on the other side of the interface (**Fig. S3**). In other regions, we observed TO layers with alternate polarity in proximity to chlorite layers (**Fig. S4**). The inheritance of the octahedral tilts in the original TO layers occurs in all these interesting structures.

**Cation Repartitioning During Chloritization**

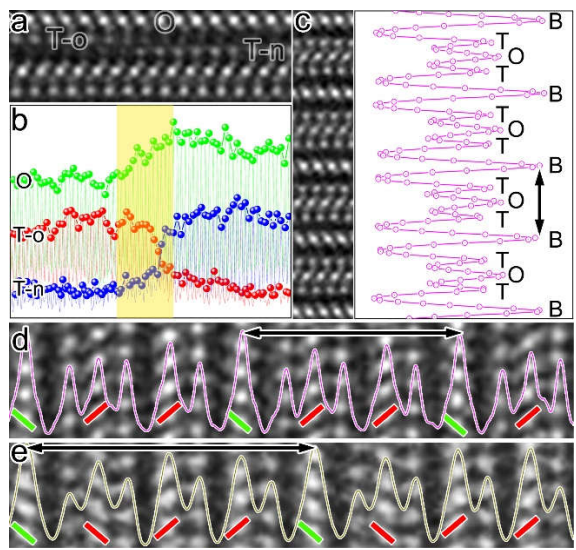

**Fig. 5. Chemical redistribution during the transformation**. (a) HAADF–STEM image of a transition region. (b) Intensity profile along the planes labeled as O, T-o and T-n. The image used for intensity analysis is shown in **Fig. S5**. The yellow region is the transition zone. (c) Intensity profile of the formed chlorite along the vertical direction of the image on the left. Intensity profile of serpentine-chlorite interstratification (d) $S_1C_1$ and (e) $S_2C_1$. Images used for intensity analysis are inserted in the background. Red and green lines illustrate the tilt of the octahedra in TOT and TO layers and brucite-like interlayers, respectively. Black arrows in (c–e) mark the periodicity for Cr-enriched layers.

In the lizardite-to-chlorite transition zones, the migration of tetrahedral cations from T-o to T-n sites can also be revealed from the atomic column intensity profiles (red and blue curves in **Fig. 5a**). Surprisingly, the intensities of the columns in O-sheet also increase throughout the transition (green curve **Fig. 5b**) suggesting that high-Z cations Cr and Fe in chlorite are not homogeneously distributed in the octahedra, as they are in lizardite layers, but become enriched in the B-interlayers (**Fig. 5c**). This also is the case for $S_1C_1$ and $S_2C_1$ (**Figs. 5d,e**).

Observation of a single chlorite layer in serpentine shows that the extra Cr and Fe in the B-interlayers originate from the O-sheets of the TOT involved in the transformation. In **Fig. 6a**, the octahedral cation columns in B-interlayers have a higher intensity than those in the lizardite layers (O in black) while the octahedral cation columns in TOT layers (O in red) have a lower intensity than those in the lizardite layers. A semi-quantitative analysis of the HAADF–STEM images using image simulation was performed. Five models were used for image simulations. In model 1, 2, 3, 4 and 5, 50%, 37.5%, 25%, 12.5% and 0% of Cr and Fe are in the octahedra in the TOT layer of chlorite. It is important to note that our inability to well resolve some oxygen atoms leads to a broader than expected peak that stems from the partially overlapping between octahedral cations and oxygen columns (green arrows) in profiles measured along the direction indicated by the red arrow in **Fig. 6c**. The best match of peak intensities between model 5 and experimental profiles leads us to conclude that essentially all Cr and Fe reside in the B-interlayers. Thus, we conclude that Cr and Fe are completely repartitioned from the octahedra sheets of lizardite to B-interlayers during the transformation.

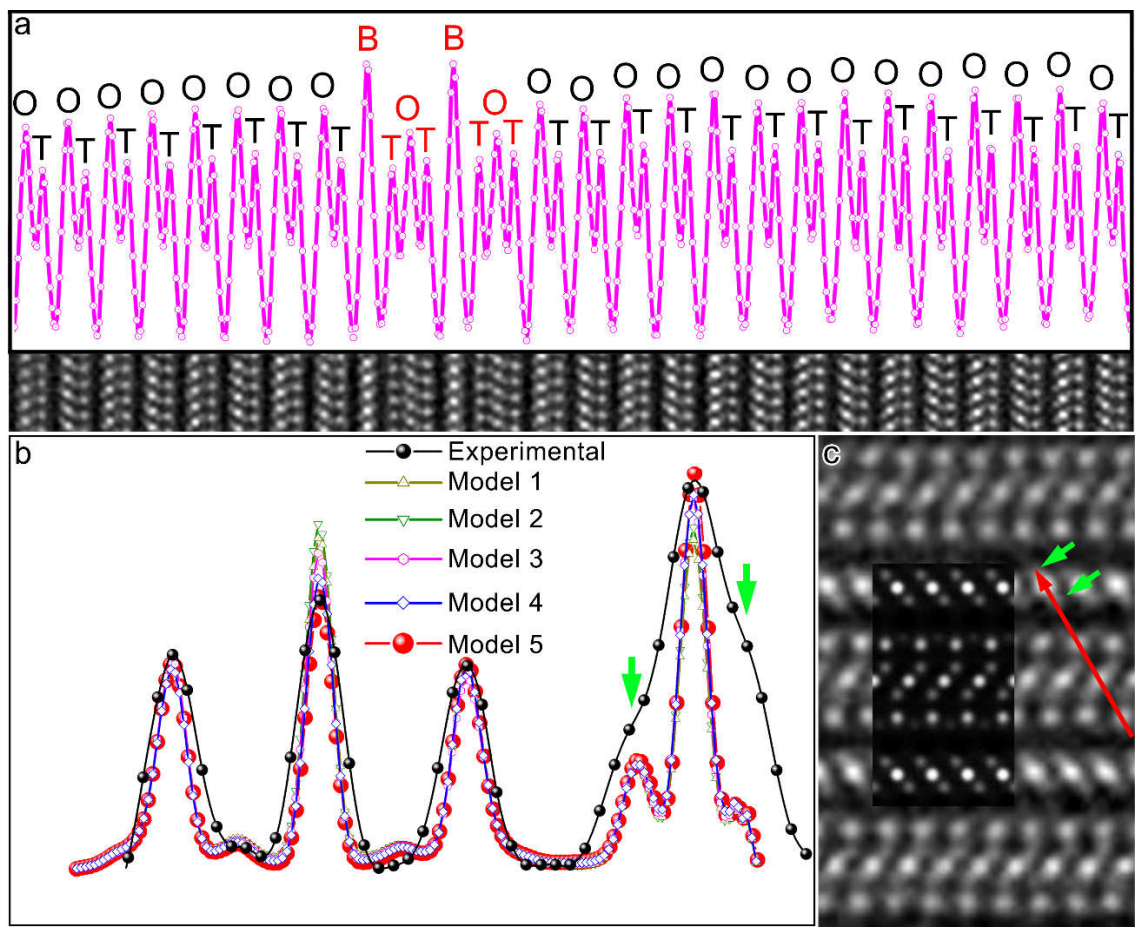

**Fig. 6. Repartitioning of high-Z cations**. (a) Intensity profile across chlorite and lizardite regions. T, O and B represent tetrahedral sheets, octahedral sheets and brucite-like interlayers, respectively. The image used to generate the intensity profile is provided below. (b) Line profiles of experimental and simulated images along the atomic columns in the direction illustrated by the red arrow in (c). Green arrows in (b) denote the shoulder peaks that deviate from the predicted intensity because they arise from oxygen columns, marked by green arrows in (c), that were not well resolved experimentally. (c) HAADF–STEM image of chlorite. An image simulated using the cation distribution of model 5 superimposed on the experimental image.

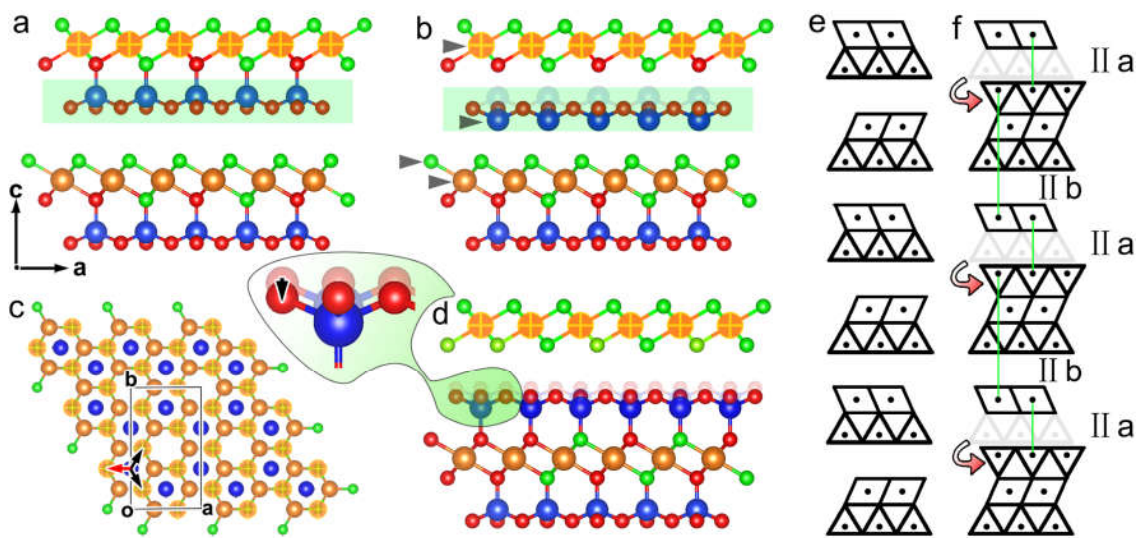

**Fig. 7. Pathway of the Serpentine–Chlorite Transformation**. (a) Projection of two TO layers along [010]. Yellow, blue, red and green balls denote the Mg, Si, oxygen and hydroxyl oxygen atoms respectively. Mg cations in the upper TO layer are highlighted by crosses. For simplicity, only the ideal model with Mg as octahedral and Si as tetrahedral cations is considered here. The structural change occurs conceptually as follows. First, Si atoms migrate ~1.2 Å downward along the *c*-axis from original positions, represented by shaded balls in (b). Note that after migration, displacement relative to the apical O below is required for Si to occupy a new tetrahedral site. Four atomic planes marked by arrows in (b) are used for the [001] projection in (c). The new tetrahedral site can be formed if the green-shaded Si–$O_3$ block in (b) is shifted by the red arrow in (c). Equivalent shifts, indicated by the black arrows, would also reestablish tetrahedral sites. This is accompanied by 0.5 Å downward shift of Si–$O_3$ block from its green-shaded position in (b), as shown in the [010] projection in (d). This displacement is shown by an arrow in the enlarged inset in (d). (e) and (f) show simplified sketches of lizardite TO layers (e) and their transformation to chlorite (f). Every second TO layer must be reconfigured to form IIab chlorite.

## DISCUSSION

*Formation of individual IIa layers and periodic IIab chlorite layers*

We have shown that the transformation of serpentine to chlorite involves two processes: the restructuring of TO layers and the repartitioning of octahedral cations. **Fig. 7** illustrates the steps by which TO-layers are restructured. In the first step, there is collective migration of tetrahedral cations and associated oxygen atoms (block indicated by green shading in **Fig. 7a**). This involves the migration of Si cations downward along the *c*-axis to flip the Si–$O_3$ block on the top of the underlying TO layer (**Fig. 7b**). A mandatory shift of -*a*/3 (or equivalent vectors, **Fig. 7c**) is required to establish new tetrahedral sites. Because Si cations form new covalent bonds with the O-sheet, the migration must be accompanied by the relocation of hydrogen atoms from the hydroxyl sheet to the new brucite-like interlayer. The final configuration is C-IIa chlorite single layer (**Fig. 7d**).

Due to the remarkable resolution achieved in low-dose HAADF–STEM images, tetrahedral cations both before and after the downward shift of Si–$O_3$ block in C-Dis were resolved (circled columns in **Fig. 2d**). The displacements of Si cations predicted by the ideal structural model in **Fig. 7** are 1.7 Å along -*c* and 0.9 Å along -*a*, perfectly agreeing with the measured displacement, 1.8 Å and 1.0 Å in **Fig. 3a**. In agreement with the observation in **Fig. 3**, the octahedra in the TOT layer and B-interlayer in **Fig. 7d** clearly tilt towards the opposite directions along the *a*-axis. If every second TO layer is transformed, it creates regular multiple chlorite layers in IIab configuration (**Fig. 7e,f**).

No non-hydrogen-bonded lizardite precursors were observed in the chloritization of lizardite in our experiments and non-hydrogen-bonded structures are not required as precursor, intermediate or final structures in any pathway observed in this study. Relative to the model proposed in Ref. 18, this work confirms that a shift of inverted T-sheet by -*a*/3 is necessary but shows that it does not require a tilt change of O-sheet in the TOT layers.

The shift of T-sheet in the present model predicts that a |*a*/3| gap, referred to as "void" in Ref. 21, is generated at the reaction interface between the transformed and untransformed structures. This atomic-scale disconnection was not observed in our experiments, but does not invalidate the

model because the varied width of transition zones (grey regions in **Fig. 2a**) shows that the reaction interfaces are inclined to viewing direction [010].

*Formation of other chlorite polytypes and interstratified structures*

The model proposed in **Fig. 7** starts from TO layers stacked without interlayer shifts along the *b*-axis. However, in Cr-lizardite, we detected many nonperiodic or semi-random shifts along the *b*-axis (**Fig. 1b,c**). Analysis of all the possible shifts along the *b*-axis indicates that the transformation scenario in **Fig. 7** applies in all cases. For lizardite with much less frequently observed interlayer shifts along the *a*-axis, a similar transformation process results in C-IIb and IIbb configurations (**Supplementary Notes** and **Fig. S6**). If the transformation is accompanied by a tilt change of the O-sheet in TOT layers, lizardite layers can be transformed into Ibb chlorite.[18] For all the chloritization of lizardite layers, no difference in the interlayer distance could be observed before and after the transformation (**Figs. 3, 4, Figs. S2, S3, S6**), signifying that the transformation is isovolumetric. Different chlorite polytypes are widely reported, and are considered to be the result of subtle differences in thermodynamic stability[45]. However, here we provide evidence that suggests that the structure of the serpentine mineral precursor can be the determining factor.

Ordered transformations that occurred in long-period lizardite with complex tilt patterns give rise to ordered serpentine-chlorite interstratifications (**Fig. 4**, **Figs. S2, S7, S8**). The out-of-phase boundaries (**Fig. S3**) and reversed polarity of serpentine layers in proximity to newly formed chlorite layers (**Fig. S4**) likely form when transformations are initiated in different regions. In other words, they are predictable consequences of the proposed solid-state transformation mechanism.

*Charge redistribution*

The enrichment of Cr and Fe in B-interlayers in chlorite and serpentine-chlorite interstratified structures results in one-dimension chemical ordering along the *c*-axis, with a periodicity of ~14.9 Å (**Fig. 5c**), ~22.6 Å (**Fig. 5d**) and ~30.0 Å (**Fig. 5e**), respectively. It has been found that Cr[46] and Fe[47] in B-interlayers preferably occupy one-half of the octahedral sites that are furthest from the tetrahedral cations, which may be in part due to the higher charges on Cr and Fe compared to Mg. The only chemical ordering in our lizardite samples was one dimensional, along the *c*-axis, possibly because the concentrations of these elements in the B-interlayers were too low for ordering to be detectable. If the randomly distributed pattern of Cr and Fe in lizardite was inherited by chlorite, the net charge on the B-interlayer would be +0.351. However, our analyses indicate that all Cr and Fe relocate to the octahedral sheet of the B-interlayer. This would result in a net charge of +0.522 on the B-interlayer. The increased charge of B-interlayer probably stabilizes the chlorite structure[46,48] by substantially increasing the interlayer binding energy, as occurs in mica.

*Destabilization mechanism of TO*

Solid-state transformations can be displacive, like martensitic transformations, or diffusive, as in the case for spinodal decomposition.[49] The serpentine-chlorite transformation involves the synergetic displacement of cations and anions, which could be regarded as a shuffle-dominated displacive process, and chemical redistribution that is diffusion-based. The transformation starts from the destabilization of lizardite TO layers. It has been well established that the lateral mismatch between T- and O-sheet dimensions and the corresponding misfit relief mechanisms distinguish the serpentine minerals,[50,51] lizardite, chrysotile and antigorite. Lizardite utilizes the coupled isomorphous substitution, Al, Cr and Fe for Mg in octahedral and Al for Si in tetrahedral sites to alleviate the misfit. Even so, misfit still exists in lizardite, as O-sheets are ~3% larger than T-sheets.[51] It is reported that O-sheets have a larger nominal thermal expansion coefficient[52] and higher compressibility[53] than T-sheets. The temperature-pressure conditions in geologic settings such as subduction zones might exacerbate the size misfit and destabilize the TO structures. Accommodation of increased size misfit can be achieved by the addition of the second T-sheet to the TO layer upon the formation of the TOT layer in the serpentine-chlorite transformation. Thermal energy also likely drives restructuring by enabling the synergetic short-range (~2 Å) migration of cations and anions, and the longer-range (~7 Å) redistribution of Cr, Fe and H.

## Conclusion

Two fundamental aspects of the isovolumetric serpentine-chlorite transformation, TO-layer restructuring and cation repartitioning, were unraveled using low-dose HAADF–STEM imaging. The tetrahedral cations collectively migrate from the destabilized TO layer to a neighboring TO layer together with the associated oxygen atoms, transforming the two TO layers into one B-interlayer and one TOT layer. Essentially all the high-Z octahedral cations in the TOT layer are repartitioned into the B-interlayer, increasing the charge on the B-interlayer. This probably contributes to the stabilization of the chlorite structure. Our work demonstrates the utility of HAADF–STEM for the study of mineral structures, microstructures and reaction mechanisms involving beam-sensitive geo-materials at the atomic scale with chemical information.

## Acknowledgments

This work was supported by the Chemical Sciences, Geosciences, and Biosciences Division, Office of Science, Office of Basic Energy Sciences, U.S. Department of Energy under award number DE-AC02-05CH11231. Work at the Molecular Foundry was supported by the Office of Science, Office of Basic Energy Sciences, of the U.S. Department of Energy under contract no. DE-AC02-05CH11231.